\documentclass[aps,twocolumn,showpacs]{revtex4}
\usepackage{graphicx}
\usepackage{epsfig}

\begin{document}

\title{Quantum state reconstruction with binary detectors}

\author{D. Mogilevtsev},
\email{dmitri@loqnl.ufal.br}

\affiliation{Institute of Physics, Belarus National Academy of
Sciences, F.Skarina Ave. 68, Minsk 220072 Belarus; \\
Departamento de F\'{\i}sica, Universidade Federal de Alagoas
Cidade Universit\'{a}ria, 57072-970, Macei\'{o}, AL, Brazil}

\begin{abstract}
I present a simple and robust method of quantum state
reconstruction using non-ideal detectors able to distinguish only
between presence and absence of photons. Using the scheme, one is
able to determine a value of Wigner function in any given point on
the phase plane using expectation-maximization estimation
technique.

\end{abstract}


\maketitle

Development of effective and robust methods of quantum state
reconstruction is a task of crucial importance for quantum optics
and informatics. One needs such methods to verify the preparation
of states, to analyze changes occurring in the process of dynamics
and to infer information about processes causing such a dynamics,
to estimate an influence of decoherence and noise-induced errors,
to improve measurement procedures and characterize quantum
devices.

For existing schemes of quantum state reconstruction losses are
the major obstacle. In the real experiment they are unavoidable;
detectors which one has to use to collect the set of data
necessary for the reconstruction are not ideal. Presence of losses
poses a limit on the possibility of reconstruction. For example,
in quantum tomography \cite{{tomogr},{tomogr1}}, which is up to
date is a the most advanced and successfully realized
reconstruction method, an efficiency of detection should exceed
$50\%$ to make possible an inference of the quantum state from the
collected data. However, the very presence of losses can be turned
to advantage and used for the reconstruction purposes.

In 1998 in the work \cite{moghrad98} it was predicted, that
non-ideal binary detectors can be used for complete reconstruction
of a quantum state. A detector able to distinguish only between
presence and absence of photons is able also to provide sufficient
data for the reconstruction. This detector must be non-ideal,
since the ideal binary detectors measures only the probability to
find the signal in the vacuum state.

To perform the reconstruction one needs a set of probe states
mixed via a beam-splitter with a signal state. For the probe
coherent states were suggested. When the probe was assumed to be
the vacuum, the procedure gives an information sufficient for
inference of a photon number distribution of the quantum state. In
Inference of the photon number distribution was discussed in works
\cite{mog98}. This scheme was implemented experimentally to
realize a multichannel fiber loop detector \cite{loop}. Very
recently it was  developed further by implementing the maximal
likelihood estimation realized with help of
expectation-maximization (EM) algorithm \cite{em1}, and
demonstrated experimentally \cite{{paris_teor},{paris_exp}}. The
reconstruction procedure with help of EM algorithm was shown to be
robust with respect to imperfections of the measurement procedure
such as, for example, fluctuations in values of detector's
efficiencies. In difference with the quantum tomography
reconstruction scheme \cite{{tomogr},{tomogr1}}, such a procedure
does not impose lower limits on detector's efficiency and requires
quite a modest number of measurements to achieve a good accuracy
of the reconstruction.

Here we demonstrate how to reconstruct a quantum state using sets
of binary detectors with different efficiencies. Let us consider a
following
simple set-up: the signal state (described by the density matrix
$\rho$) is mixed on a beam-splitter with the probe coherent state
$|\beta\rangle$. Then the probability $p$ to have \textit{no}
counts simultaneously on two detectors is measured (as it can be
seen later, in fact, it is possible to use only one detector for
the reconstruction).

According to Mandel's formula, this probability is
\begin{equation}
p=\langle:\exp{\{-\nu_cc^{\dagger}c-\nu_dd^{\dagger}d\}}:\rangle,
\label{p01}
\end{equation}
where $\nu_c,\nu_d$ are efficiencies of the first and second
detectors; $c^{\dagger},c$ and $d^{\dagger},d$ are creation and
annihilation operators of output modes and $::$ denoted the normal
ordering. For simplicity we assume here, that there is no `dark
current', and in absence of the signal detectors produce no
clicks. Let us assume, that the beam-splitter transforms input
modes $a$ and $b$ in the following way
\begin{equation}
c=a\cos(\alpha)+b\sin(\alpha),\qquad
d=b\cos{(\alpha)}-a\sin{(\alpha)}. \label{tr1}
\end{equation}
Then averaging over the probe mode $b$, from Eqs. (\ref{p01}) and
(\ref{tr1}) one obtains
\begin{equation}
p=e^yTr\{:\exp{\{-{\bar\nu}(a^{\dagger}+\gamma^*)(a+\gamma)\}}:\rho\},
\label{p02}
\end{equation}
where
\begin{eqnarray}
\nonumber {\bar\nu}=\nu_c\cos^2(\alpha)+\nu_d\sin^2(\alpha), \\
\gamma=\beta(\nu_d-\nu_c)\cos(\alpha)\sin(\alpha)/{\bar\nu},
\\ \nonumber
y=-|\beta|^2\nu_c\nu_d\sin^2(2\alpha)/{\bar\nu}. \label{coef1}
\end{eqnarray}
Finally, from Eq. (\ref{p02}) one obtains
\begin{equation}
p=e^y\sum\limits_{n=0}(1-{\bar\nu})^n\langle
n|D^{\dagger}(\gamma)\rho D(\gamma)|n\rangle, \label{p03}
\end{equation}
where $D(\gamma)=\exp{\{\gamma a^{\dagger}-\gamma^*a\}}$ is a
coherent shift operator, and $|n\rangle$ denotes a Fock state of
the signal mode $a$.

Essence of the reconstruction procedure is in measurement of $p$
for different values of ${\bar\nu}$ and fixed value of the
parameter $\gamma$. Let us, for example, assume detectors
efficiencies $\nu_c,\nu_d$ to be constant, and $\nu_c\neq\nu_d$
(one of efficiencies can be set to zero; only one detector might
be used in the scheme). Then for arbitrary $\gamma$ let us mix the
signal state with the probe coherent state having the amplitude
\begin{equation}
\beta_j=2{\gamma(\nu_c\cos^2(\alpha_j)+\nu_d\sin^2(\alpha_j))\over
(\nu_d-\nu_c)\sin(2\alpha_j)}
 \label{beta}
\end{equation}
and measure a set of probabilities $p$  for different values of
the beam-splitter rotation angle $\alpha_j$. Then we have linear
positive inverse problem of finding quantities
\begin{eqnarray}
R_{n}(\gamma)=\langle n|D^{\dagger}(\gamma)\rho
D(\gamma)|n\rangle,
\end{eqnarray}
which could be solved by the EM iterative algorithm similar to the
one used in works \cite{{paris_teor},{paris_exp}} for the
reconstruction of diagonal elements of the signal state density
matrix. Besides, $R_{n}(0)$ are diagonal elements of the density
matrix of the signal.

The EM algorithm for the suggested scheme of the reconstruction is
as follows. We assume the signal density matrix to be finite in
Fock-state basis being $N\times N$, and the number of different
values of $\alpha_j$ is $M\geq N$. Then we chose an initial set of
$R^{(0)}_n(\gamma)>0$, $\forall n$, and implement the following
iterative procedure \cite{{em1},{hrad1}}:
\begin{equation}
R^{k+1}_n(\gamma)=R^{k}_n(\gamma)\sum\limits_{j=0}^{M-1}{(1-{\bar\nu}_j)^np_j^{exp}
\over f_jp_j^{(k)}},
 \label{em}
\end{equation}
where $p_j^{exp}$ is the experimentally measured frequency of
having no clicks on both detectors for a given $\alpha_j$, and
$p_j^{(k)}$ is the left-hand side of Eq. (\ref{p03}) calculated
using the result of $k$-th iteration. The weights
\begin{eqnarray}
\nonumber f_j=\sum\limits_{n=0}^{N-1}(1-{\bar\nu}_j)^n.
\end{eqnarray}
The procedure (\ref{em}) guarantees positiveness and unit sum of
the reconstructed $R_n(\gamma)$.

Finally, having reconstructed quantities $R_n(\gamma)$, it is
straightforward to find a value of the Wigner function at the
point $\gamma$ \cite{wig}:
\begin{equation}
W(\gamma^*,\gamma)={2\over\pi}\sum\limits_{n=0}^{N-1}(-1)^nR_n(\gamma).
 \label{wign}
\end{equation}

The scheme can be even more simplified if we set the efficiency of
the second detector to zero $\nu_d=0$. Then $p$ is simply the
probability to have no clicks on a single detector. We have also
$y=0$, and the parameter $\gamma$ does not depend on the
efficiency:
\begin{equation}
\gamma=-\beta \tan(\alpha).
 \label{gnew}
\end{equation}

Thus, one can determine $R_n(\gamma)$ by varying the efficiency
$\nu_c$ and keeping $\alpha$ constant without even changing the
amplitude of the probe coherent state $\beta$.

\begin{figure}[th]
\centerline{\psfig{file=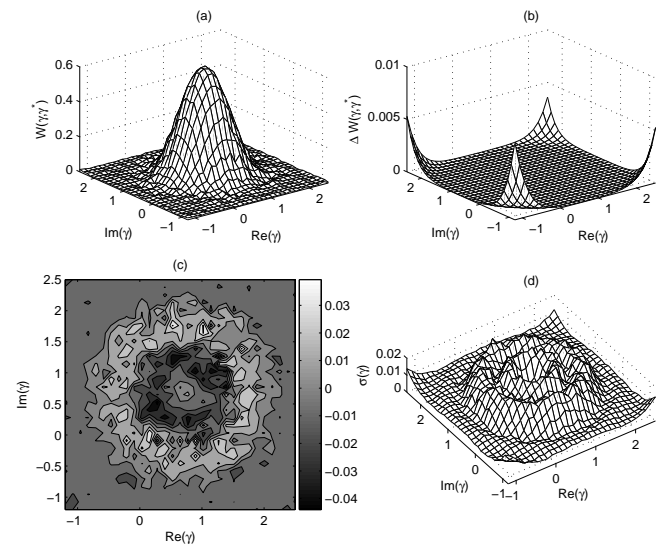,width=\linewidth}}
\vspace*{8pt} \caption{Reconstruction of the signal coherent state
$\alpha$ with the amplitude $\alpha=1$. In Figure (a) the
reconstructed Wigner function is shown; in figure (b) the
difference between the exact Wigner function and the Wigner
function of the truncated state is shown. Figure (c) shows
difference between the exact and the reconstructed Wigner
functions; in Figure (d) the variance $\sigma(\gamma,\gamma^*)$ is
depicted. For all pictures $N_r=10^4$ measurements were used for
each point on the phase plane and $N_{it}=10^3$ iteration of the
EM algorithm. The state was truncated with $N=12$; 30 different
values of the detector efficiency were taken; they were
distributed homogeneously in the interval $[0.1,0.9]$.}
\label{fig2}
\end{figure}

In Figure~\ref{fig2} the reconstruction of the Wigner function of
the signal coherent state with help of the one-detector version of
the method is illustrated.

Starting from the uniform distribution $R_n^{(0)}(\gamma)=1/N$,
very good results of the reconstruction was achieved with only a
$10^3$ iterations of the EM algorithm and $10^4$ measurements for
each point on the phase plane. As it was mentioned in the work
\cite{{paris_teor}}, the choice of $R_n^{(0)}(\gamma)\neq 0$ was
indeed not influencing much the convergence for any given point.
However, for different points on the phase plane the rate of
convergence might differ strongly. In region of more rapid change
of the Wigner function one needs more iterations and more
measurements to achieve the same precision (as it can bee seen in
Figure~\ref{fig2}(d); here the variance is smaller near the peak
of $W(\gamma,\gamma^*)$). An explanation can be easily found from
the formula (\ref{wign}): in the region of rapid change one needs
to find with high precision several comparable $R_n(\gamma)$,
whereas, for example, the behavior of the Wigner function near
$\gamma=\alpha$ is defined mostly by $R_0(\gamma)$. Also, the
precision is influenced by the truncation number $N$. Increasing
the region on the phase plane, where the Wigner function is to be
estimated, one needs also to increase $N$. In Figure~\ref{fig2}(b)
one can see the difference between the exact and truncated Wigner
functions. As a consequence (as Figures~\ref{fig2}(c) and (d)
show), in the regions when the truncation error is significant
errors of the reconstruction procedure are also increased.

For illustration of how the total error of the reconstruction
propagates, we use the average distance between values of the
exact and the reconstructed Wigner functions
\begin{equation}
\delta W={1\over
N_p}\sum\limits_{\forall\gamma}|W_{exact}(\gamma^*,\gamma)-W(\gamma^*,\gamma)|,
 \label{dwig}
\end{equation}
where the summation taken over all points on the phase plane were
the estimation was made; $N_p$ stands for the number of points on
the phase plane.
\begin{figure}[th]
\centerline{\psfig{file=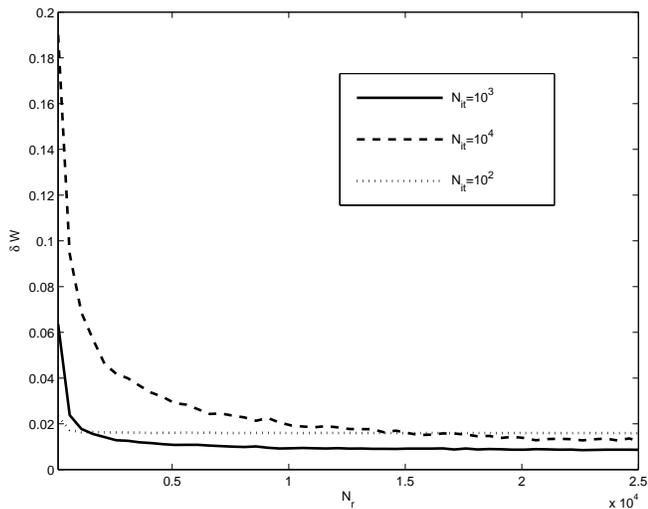,width=\linewidth}}
\vspace*{8pt} \caption{The propagation of the reconstruction error
$\delta W$(\ref{dwig}) for different numbers of iterations
$N_{it}$ in dependence on the number of experiment runs $N_r$; for
all curves $N_p=2500$ and the following region on the phase plane
was taken $Re(\gamma),Im(\gamma)\in [-1.2,2.5]$. Other parameters
are as in Figure~\ref{fig2}.} \label{fig3}
\end{figure}
It can be seen in Figure~\ref{fig3} that for smaller number
 of iterations $N_{it}$ an increase of $N_r$ leads to quicker
convergence for small number of measurements; after that the error
$\Delta W$ decreases very slowly with increasing of $N_r$. An
increase in the number of iteration leads to much slower
convergence for small $N_r$. However, with increasing of $N_r$ an
accuracy improve more rapidly; the error goes below values
achieved for smaller number of iterations. Generally,
Figure~\ref{fig3} confirms an observation made in the work
\cite{paris_teor}: for performing the reconstruction procedure it
is reasonable to use a number of iterations close to the number of
measurements, since for $N_{r}\gg N_{it}$ an accuracy improves
very slowly, and too large $N_{it}$ might even lead to increasing
of $\Delta W$.

From the practical point of view, to perform the reconstruction it
is reasonable first to estimate a photon number distribution (to
find the set of $R_n(0)$). This will provide a clue for estimating
the region of the plane sufficient for the reconstruction, and
also the truncation number necessary for the purpose.

\begin{figure}[th]
\centerline{\psfig{file=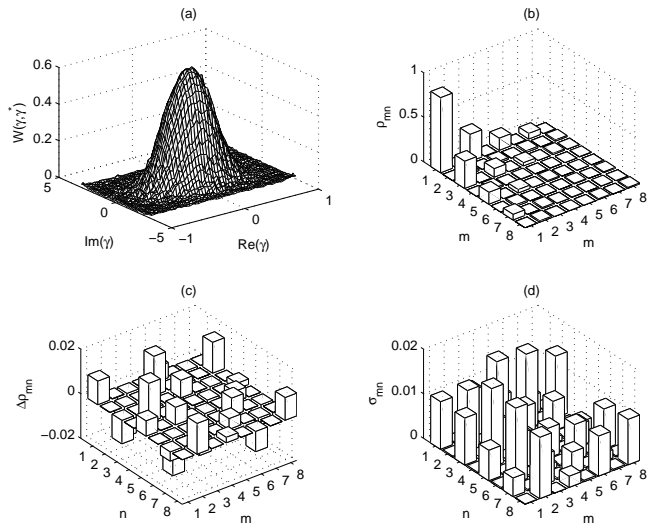,width=\linewidth}}
\vspace*{8pt} \caption{Reconstruction of the squeezed coherent
state (\ref{sq1}) with $\tanh(r)=0.5$. In Figure (a) the
reconstructed Wigner function is shown; in Figure (b) one can see
$\rho_{mn}$ obtained with help of Eq. (\ref{ro1}). In Figure (c)
the difference between exact and reconstructed $\rho_{mn}$ is
shown; Figure (d) shows the variance $\sigma_{mn}$. For all
Figures the Wigner function was found in $N_p=2500$ points; the
following region of the phase plane was used $Re(\gamma)\in
[-1.1]$, $Im(\gamma)\in [-3,3]$. Other parameters are as in
Figure~\ref{fig2}. } \label{fig4}
\end{figure}
It is possible to infer elements $\rho_{mn}$ of the signal state
density matrix in the Fock state basis  using the reconstructed
Wigner function in the following way \cite{glaub}:
\begin{equation}
\rho_{mn}=2\int d^2\gamma (-1)^nW(\gamma^*,\gamma)D_{mn}(2\gamma),
 \label{ro1}
\end{equation}
where
\begin{eqnarray}
\nonumber D_{mn}(2\gamma)=\exp{\{-2|\gamma|^2\}}\sqrt{m!n!}\times
\\ \nonumber
 \sum\limits_{l=0}^{min(m,n)}{(2\gamma)^{n-l}(-2\gamma^*)^{m-l}\over
 l!(m-l)!(n-l)!}.
\end{eqnarray}

In Figure~\ref{fig4} an example of matrix elements $\rho_{mn}$ of
the squeezed vacuum state
\begin{equation}
|r\rangle=\exp\{-r^2(a^{\dagger2}-a^2)/2\} \label{sq1}
\end{equation}
is demonstrated. One can see that even for modest number of points
on the phase plane ($50$ points along each axis) and measurements
($10^4$ per point) the accuracy of the reconstruction of
$\rho_{mn}$ is remarkable. The truncation error does not influence
this elements much, because the function $D_{mn}(2\gamma)$ is
small in the regions where the truncation error strongly
influences the reconstructed Wigner function. One can mention
here, that for reconstruction of $\rho_{mn}$ from the quantities
$R_n(\gamma)$ one does not need to find the Wigner function in the
whole region required for the integral $\int d^2\gamma
W(\gamma,\gamma^*)$ to be close to unity. It is sufficient to find
$R_{n}(\gamma)$ on $N$ circles on the phase plane \cite{wel}.
However, this approach leads to non-positive problem of inferring
$\rho_{mn}$ from the set of $R_{n}(\gamma)$.

To conclude, we have suggested and discussed a simple and robust
method of the reconstruction of the quantum state of light. For
the method one needs to use a binary detector, a  coherent state
for the probe and a beam-splitter. The reconstruction problem is
linear and positive, and solved with help of the fast and
efficient EM algorithm of the maximal likelihood estimation. With
help of the method a value of the Wigner function of the signal
state can be found in any point on the phase plane.

The author acknowledges financial support from the Brazilian
agency CNPq and the Belarussian foundation BRRFI.

\end{document}